\documentclass[10pt,twocolumn]{article}

\usepackage[margin=0.75in]{geometry}
\usepackage{times}
\usepackage{graphicx}
\usepackage{amsmath, amssymb}
\usepackage{authblk}
\usepackage{abstract}
\usepackage{caption}
\usepackage{booktabs}
\usepackage{hyperref}
\usepackage{float}
\usepackage{appendix}
\usepackage{caption}
\usepackage{subcaption}

\title{\bfseries Computational Analysis of Disease Progression in Pediatric Pulmonary Arterial Hypertension}

\author[1,2]{Omar Said}
\author[1]{Christopher Tossas-Betancourt}
\author[3]{Mary K. Olive}
\author[3]{Jimmy C. Lu}
\author[3]{Adam Dorfman}
\author[1,4]{C. Alberto Figueroa}

\affil[1]{Department of Biomedical Engineering, University of Michigan, Ann Arbor, MI, USA}
\affil[2]{Department of Computer Science, University of Michigan, Ann Arbor, MI, USA}
\affil[3]{Department of Pediatrics, Division of Pediatric Cardiology, University of Michigan, Ann Arbor, MI, USA}
\affil[4]{Department of Surgery, University of Michigan, Ann Arbor, MI, USA}

\date{}

\begin{document}
\twocolumn[
\maketitle
\vspace{-0.75em}
\begin{abstract}
Pulmonary arterial hypertension (PAH) is a progressive cardiopulmonary disease that leads to increased pulmonary pressures, vascular remodeling, and eventual right ventricular (RV) failure. Pediatric PAH remains understudied due to limited data and the lack of targeted diagnostic and therapeutic strategies. In this study, we developed and calibrated multi-scale, patient-specific cardiovascular models for four pediatric PAH patients using longitudinal MRI and catheterization data collected approximately two years apart. Using the CRIMSON simulation framework, we coupled three-dimensional fluid-structure interaction (FSI) models of the pulmonary arteries with zero-dimensional (0D) lumped-parameter heart and Windkessel models to simulate patient hemodynamics. An automated Python-based optimizer was developed to calibrate boundary conditions by minimizing discrepancies between simulated and clinical metrics, reducing calibration time from weeks to days. Model-derived metrics such as arterial stiffness, pulse wave velocity, resistance, and compliance were found to align with clinical indicators of disease severity and progression. Our findings demonstrate that computational modeling can non-invasively capture patient-specific hemodynamic adaptation over time, offering a promising tool for monitoring pediatric PAH and informing future treatment strategies.
\end{abstract}
\vspace{1em}
]

\section*{Introduction}

Pulmonary Arterial Hypertension (PAH) is a severe, progressive cardiopulmonary disease characterized
by elevated blood pressure in the pulmonary arteries. PAH patients experience structural and functional
changes in the pulmonary vasculature, resulting in hemodynamic alterations that can eventually lead to
right ventricular (RV) failure$^{1}$. Changes in hemodynamics, such as elevated pressure, can trigger
adaptations in the vessel walls, such as stiffening and thickening. Such structural adaptations can lead to
further increases in pressure, creating a positive feedback loop that plays a critical role in the progression
of PAH, as seen in Figure~1$^{2}$. Ventricular-arterial interactions play an important role in the progression of
PAH, where increases in resistance and decreases in compliance of the pulmonary circulation led to
structural remodeling and increased contractility of the RV, in an attempt to maintain normal cardiac
outputs shown in Figure~2$^{3}$. RV contractility can increase up to four- to five-fold compared to healthy
levels in an effort to overcome elevated pulmonary pressures. However, this compensatory mechanism
has a limit—once exceeded, it results in ``uncoupling'' between the RV and pulmonary arteries. This
uncoupling leads to a decline in RV stroke volume and ejection fraction, ultimately progressing to
decompensated RV failure. As such, identifying reliable metrics to characterize disease progression is
critical for improving PAH prognosis.

\begin{figure}[H]
    \centering
    \includegraphics[width=\columnwidth]{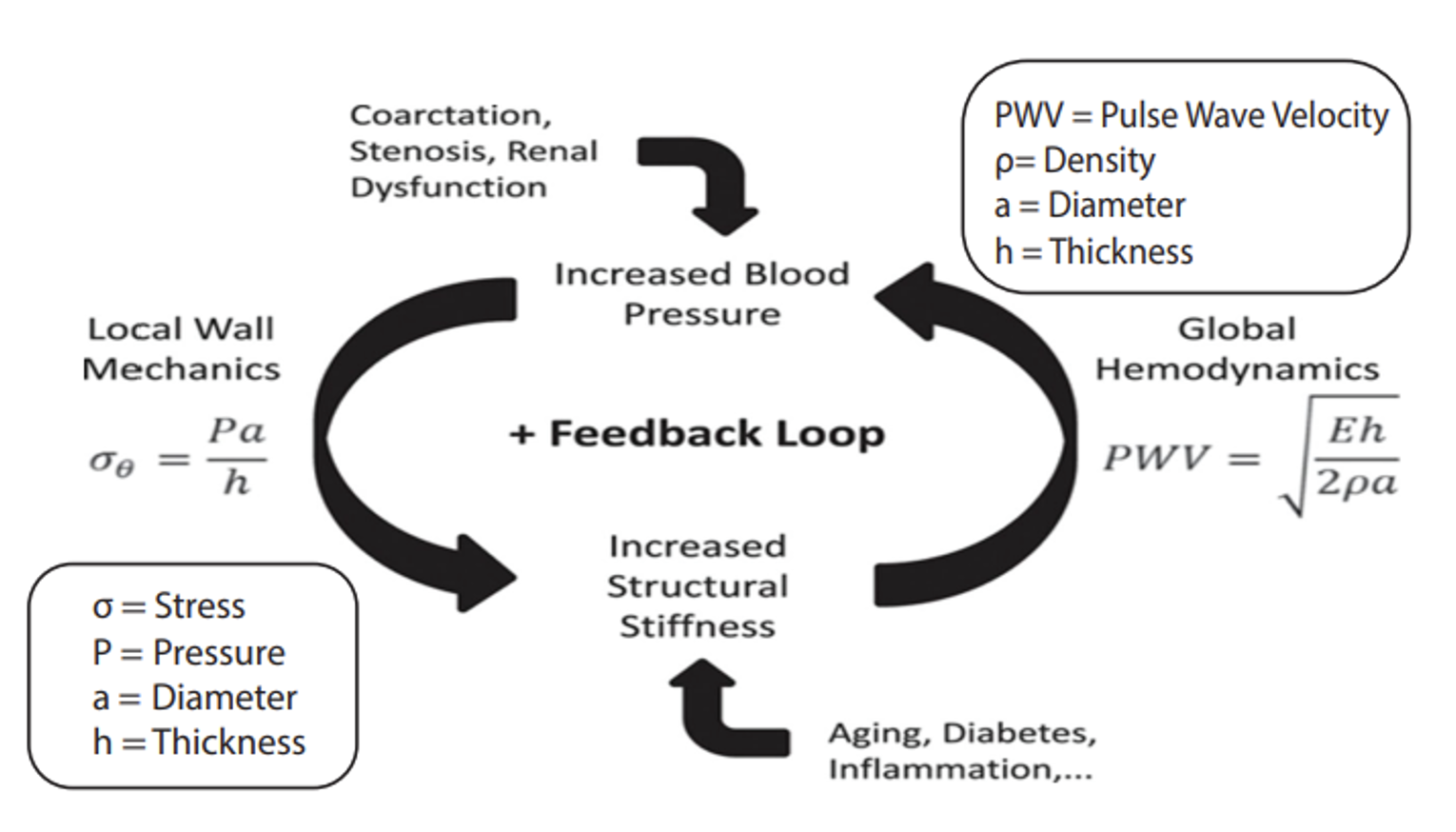}
    \caption{Schematic of the vascular stiffening feedback loop. Increased blood pressure and structural stiffness reinforce each other through local wall mechanics and global hemodynamics, influenced by aging, disease, and vascular remodeling. Equations relate wall stress and pulse wave velocity (PWV) to arterial geometry and material properties.}
    \label{fig:feedback_loop}
\end{figure}

Pediatric PAH is understudied especially when it comes to diagnosis the disease compared to adult PAH
and that is primarily due to the lack of data and clinical trials in the pediatric population. Management of
pediatric PAH is primarily guided by clinical experience and evidence derived from adult studies, due to a
significant lack of pediatric-specific research. Key challenges include the limited evaluation of diagnostic
and therapeutic strategies in children, uncertainty around drug toxicity and optimal dosing, and the
absence of validated clinical trial endpoints tailored to the pediatric population. Additionally, there is a
need for well-defined treatment targets to support goal-directed care in children. These gaps present major
obstacles for developing effective and evidence-based therapies for pediatric PAH$^{5}$.

\begin{figure}[H]
    \centering
    \includegraphics[width=\columnwidth]{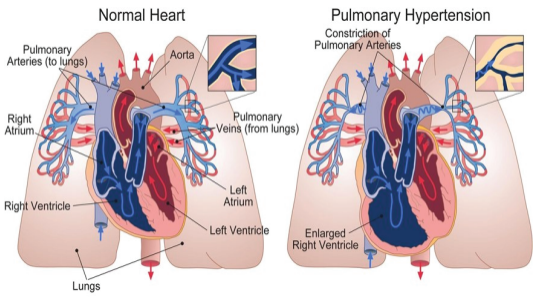}
    \caption{Comparison between a normal heart and a heart affected by pulmonary hypertension. Note the constriction of the pulmonary arteries and right ventricular enlargement in the diseased state.}
    \label{fig:heart_comparison}
\end{figure}

This study builds from a previous study$^{6}$, where our group reported methods to calibrating said
computational models, and correlated data-derived and model-derived metrics to disease severity, using
data from a single timepoint by constructing 3D bi-directional model of the cardiopulmonary system,
incorporating both ventricular and arterial hemodynamics$^{3}$.

We will investigate PAH progression in children by leveraging longitudinal patient data to build patient-
specific computational hemodynamic models. Here, we aim to leverage a prospective longitudinal patient
dataset and computational models to yield patient-specific metrics at two timepoints separated by $\sim$2
years. This will enable us to evaluate correlations between metrics and PAH progression.

Ultimately, our work aims are to: 1) develop and calibrate multi-scale closed-loop models of the
cardiopulmonary circulation in PAH patients and that would help validate the methodology used on
previous study from our research group, and 2) Use a combination of clinical data and model-derived
hemodynamic parameters to categorize patients based on the extent of their disease progression. We will
utilize a multi-scale modeling framework that couples three-dimensional, image-based representations of
the pulmonary arteries with lower-dimensional (0D) lumped-parameter networks of the remaining
cardiopulmonary circulation. This approach captures critical arterial interactions that underlie disease
progression, particularly relevant in pediatric patients where standard adult-based metrics often fail to
account for differences in cardiovascular structure and function. Our longitudinal dataset spans
hemodynamic, imaging, and clinical measurements, facilitating the calibration of patient-specific
boundary conditions and material properties at each timepoint. By incorporating precise morphological
and biomechanical information into our models, we can more accurately capture hemodynamics and
vessel wall adaptation.

In summary, this study builds upon prior multiscale computational modeling work in pediatric PAH that established correlations between model-derived hemodynamic metrics and disease severity at a single timepoint. Here, we extend this framework to a longitudinal cohort, focusing on within-patient changes over time and introducing an automated calibration pipeline to enable scalable, reproducible modeling across follow-up studies$^{6}$.

\section*{Methods}

\textbf{Clinical Data:} Clinical data on anatomy, flow, and pressure were prospectively acquired using MRI and
pressure catheterizations in four pediatric PAH patients (ClinicalTrials.gov ID No. NCT03564522). Two
separate datasets (baseline and follow-up) were captured for each patient, separated approximately 2
years, to study disease progression.

\textbf{Clinical Metrics:} In this study, 39 clinical metrics were analyzed to evaluate disease progression,
spanning three main categories: patient-specific characteristics (such as age, body size, and surface area),
imaging-based cardiac function metrics obtained from MRI (including ventricular volumes, mass, and
cardiac output), and invasive hemodynamic data acquired through catheterization (such as pulmonary
pressures and vascular resistance), as shown in Table 1 in the Appendix.

\textbf{Imaging and Hemodynamic Assessment:} Cardiac MRI was performed using a 1.5T scanner (Achieva
or Ingenia; Philips, Netherlands). Three-dimensional vascular anatomy was captured using a gated 3D
steady-state free precession (SSFP) sequence, while time-resolved blood flow and luminal area
measurements were obtained through gated phase-contrast MRI (PC-MRI) across 40 cardiac phases at
five anatomical sites: the ascending aorta (AAo), descending thoracic aorta (DTA), main pulmonary
artery (MPA), left pulmonary artery (LPA), and right pulmonary artery (RPA).

In parallel, all patients underwent right heart catheterization to collect invasive pressure data from the
right atrium, right ventricle (RV), pulmonary arteries, and pulmonary capillary wedge. Femoral arterial
access allowed systemic pressure monitoring and blood gas analysis. In select cases, left heart
catheterization was performed to assess LV and aortic pressures. Pulmonary vascular resistance index
(PVRi) was calculated using pressure gradients and cardiac index derived by Fick and thermodilution
methods.

\textbf{Patient Demographic:} Four PAH patients (age: 15.0 ± 4.4 years; range 8 – 21 years) underwent
catheterization and MRI examinations. Three patients were classified as WHO-FC I and one as WHO-FC
II. Two patients were on dual PAH therapy, one on triple PAH therapy, and one on quad PAH therapy.

\textbf{High-resolution Arterial Model:} Patient-specific cardiovascular models were developed using the open-
source simulation platform CRIMSON$^{7}$. For each patient, closed-loop 3D fluid-structure interaction (FSI)
models were constructed at two timepoints using detailed anatomical and functional imaging data. These
3D models were coupled with zero-dimensional (0D) lumped-parameter representations of the heart (H)
and the distal pulmonary and systemic circulations, modeled using three-element Windkessel models
(W)$^{8}$. The Windkessel models captured the resistance and compliance of peripheral vasculature, while the
heart model simulated atrial and ventricular dynamics, including valve function. CRIMSON generated
smooth NURBS-based vessel geometries, which were meshed into linear tetrahedral elements for
numerical simulation. Data- and model-derived metrics were analyzed to assess longitudinal changes in
patient hemodynamics. The combined finite element mesh size for both the aortic and pulmonary models
ranged from 1,414,936 to 1,890,093 elements. Computational models were used to yield quantitative
metrics such as pressure, flow, pulse wave velocity, pulmonary resistance and capacitance, and arterial
stiffness.

Clinical hemodynamic data informed arterial wall properties (including linearized stiffness and thickness)
as well as inflow and outflow boundary conditions. The arterial wall was represented as a linear elastic
membrane characterized by spatially varying isotropic stiffness and thickness$^{9}$. Linearized stiffness was
derived using luminal area and pressure data, defined as follows$^{10,11}$:
\begin{equation}
E = \frac{1.5 \cdot \Delta P \cdot R_i^2 \cdot R_o}{(R_o^2 - R_i^2) \cdot \Delta R}
\end{equation}

where R$_i$ and R$_o$ represent the diastolic luminal radius and the outer vessel radius, respectively.
$\Delta R = R_{systolic} - R_{diastolic}$ is the change in lumen radius, and $\Delta P = P_{systolic} -
P_{diastolic}$ represents the pulse pressure. A wall thickness-to-vessel radius ratio of 15\% was applied
to the large systemic$^{12}$ and pulmonary$^{13}$ arteries. Linearized stiffness was evaluated at five
locations—AAo, DTA, MPA, LPA, and RPA—where arterial wall deformation was estimated via PC-MRI.
Stiffness values were linearly interpolated along the vessel centerline, and branch stiffness was assigned
based on the nearest large arterial segment.

After defining the parameters for both the 3D and 0D compartments of the model, multi-scale FSI simulations were performed using the CRIMSON flow solver, solving the
Navier-Stokes equations for an incompressible Newtonian fluid$^{9,14,15}$. All simulations utilized a time
step size of 0.1 ms. Blood was modeled as an incompressible Newtonian fluid with a density of
$\rho = 0.00106$ g/mm$^3$ and viscosity of $\mu = 0.004$ g/mm$\cdot$s. Simulations were run until flow
and pressure fields achieved cycle-to-cycle periodicity.

A 0D lumped-parameter heart model (H) was established utilizing CRIMSON’s
Netlist Editor Boundary Condition Toolbox$^{16}$, as shown in Figure~3. The lumped-parameter heart model
used in this work, developed by a group in Stanford$^{17}$, effectively describes the interaction between
cardiac function and arterial characteristics. The model was selected due to its integration within the
CRIMSON solver and its extensive application in 3D hemodynamic simulations$^{18,19,20}$. The aortic and
pulmonary artery roots were represented capacitors (C$_R$). The mitral and tricuspid valves were
represented through diodes (D$_{v1}$) paired with inductors (L$_{v1}$) using predefined values, while
dynamically controlled resistors (R$_{v2}$) and inductors (L$_{v2}$) were used to simulate the aortic and
pulmonary valves$^{21,22}$. A time-varying pressure volume chamber representing ventricular elastance
(E$_v$(t)) and a dynamic source resistance (R$_s$) represented the LV and RV contractility.

\begin{figure}[H]
    \centering
    \includegraphics[width=\columnwidth]{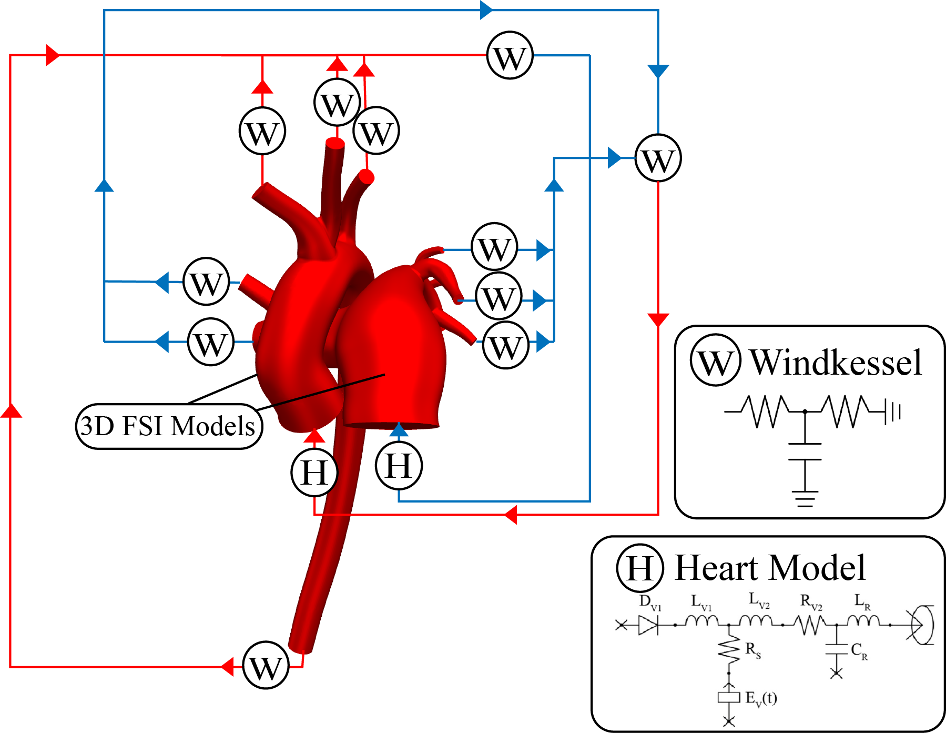}
    \caption{Schematic of the multi-scale cardiovascular simulation framework. The 3D FSI model of the heart and vasculature is coupled with lumped-parameter Windkessel models (W) at outlets and a 0D heart model (H) at the inlet to capture global hemodynamics.}
    \label{fig:multiscale_framework}
\end{figure}

\textbf{Boundary Conditions:} The design and calibration of boundary conditions were carried out through a
three-stage process involving the construction of lumped-parameter circuits, iterative parameter
adjustments, and fine-tuning of ventricular elastance and volume waveforms, as shown in Figure 4.
Specifically, boundary conditions were progressively developed in three distinct stages$^{16}$: Stage 1
employed an open-loop arterial model driven by prescribed flows at the aorta and main pulmonary artery.
Stage 2 introduced a 0D heart model into the open-loop arterial setup. Finally, Stage 3 implemented a
fully closed-loop arterial model integrated with the 0D heart model. At each stage, parameters were
iteratively optimized until the simulated flow and pressure results closely matched clinical
measurements$^{23}$. Once calibrated, these parameters were subsequently applied to the lumped-parameter
circuits in the following stage.

\begin{figure*}[t]
    \centering
    \includegraphics[width=\textwidth]{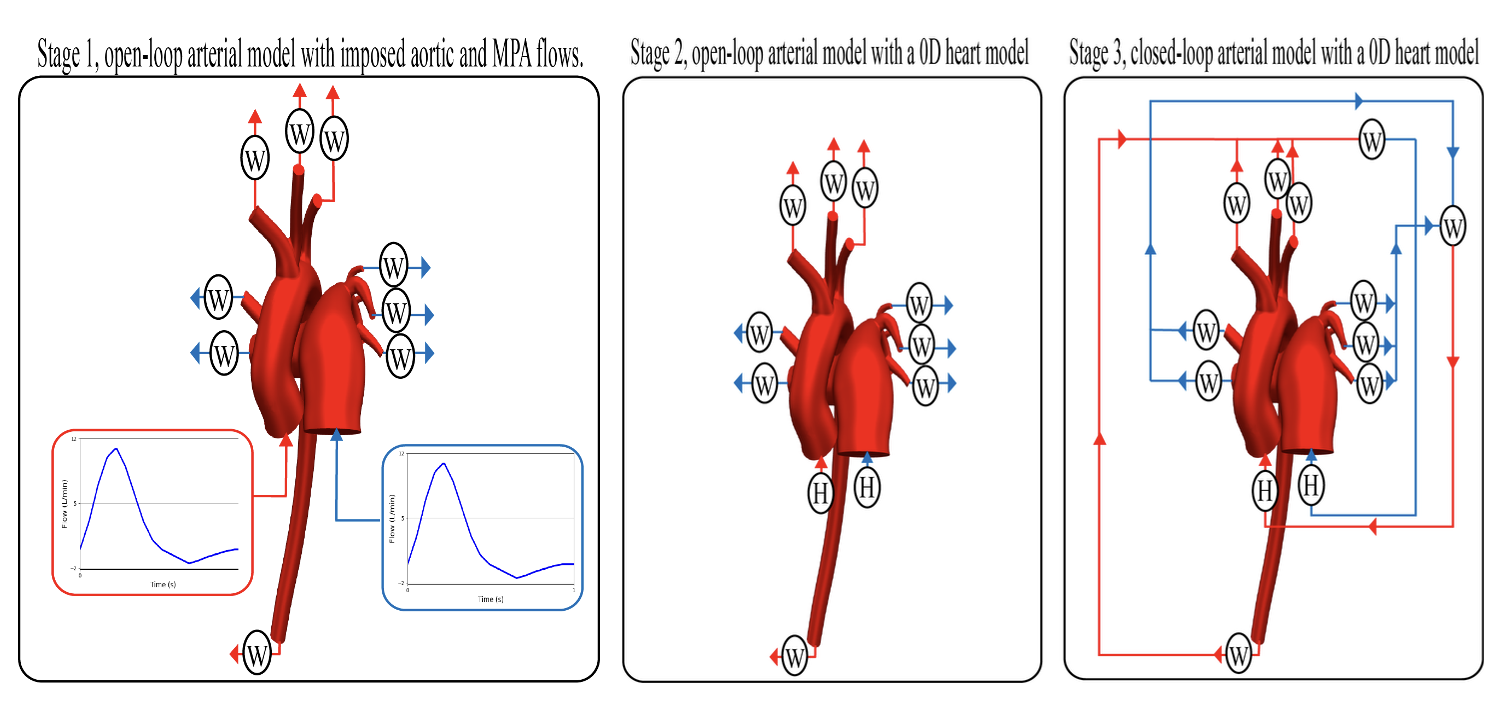}
    \caption{Boundary condition calibration workflow across three stages. 
    \textbf{Stage 1} uses an open-loop arterial model with imposed aortic and MPA flows. 
    \textbf{Stage 2} integrates a 0D heart model (H) while maintaining open-loop architecture. 
    \textbf{Stage 3} forms a fully closed-loop system coupling the arterial model with the heart model and Windkessel outlets.}
    \label{fig:boundary_conditions}
\end{figure*}

\textbf{Tuning lumped parameters:} Lumped-parameter values were determined using a fixed-point iteration
approach$^{23,24}$. The iterative calibration method utilizes the one-dimensional (1D) arterial models to
efficiently estimate outflow boundary condition parameters suitable for three-dimensional (3D), patient-
specific arterial simulations$^{23}$. Based on this approach, we derived the iterative formulas below to align
simulation outputs with clinical hemodynamic measurements.

\textbf{Stage 1, Open-loop arterial model with imposed aortic and MPA flow:} 3-element Windkessel models
were used to represent the resistance and compliance of the distal vascular bed. Windkessel resistances $R_i$
and compliances $C_i$ for each outlet branch $j$ were iteratively tuned using:
\begin{equation}
R_{j}^{n+1} = R_{j}^{n}\frac{R_{T}^{n+1}}{R_{T}^{n}}, \;\;\;\;\;
C_{j}^{n+1} = C_{j}^{n}\frac{C_{T}^{n+1}}{C_{T}^{n}}
\end{equation}
where the total arterial resistance $R_T$ and total arterial compliance $C_T$ were iteratively estimated as:
\begin{equation}
R_{T}^{n+1} = R_{T}^{n} + \frac{P_{\textit{mean}} - P_{\textit{mean}}^{n}}{Q_{\textit{mean}}^{n}}
\end{equation}
\begin{equation}
C_{T}^{n+1} = C_{T}^{n}\frac{P_{\textit{pulse}}}{P_{\textit{pulse}}^{n}}
\end{equation}
where n is the iteration counter. Simulated ($P_{\textit{mean}}^{n}$, $P_{\textit{pulse}}^{n}$) and measured pressures
($P_{\textit{mean}}$, $P_{\textit{pulse}}$) were compared at the DTA and MPA.

\textbf{Stage 2, open-loop arterial model with a 0D heart model:} Initial nodal pressures of the lumped-
parameter heart models and Windkessel models for each branch were iteratively tuned:
\begin{equation}
P_{\textit{initial}}^{n+1} = P_{\textit{initial}}^{n} \cdot \textit{mean}\left(\frac{Q_{\textit{mean}}}{Q_{\textit{mean}}^{n}},\frac{P_{\textit{mean}}}{P_{\textit{mean}}^{n}}\right)
\end{equation}
Simulated ($Q_{\textit{mean}}^{n}$) and measured flow rates ($Q_{\textit{mean}}$) were compared at the AAo and MPA.

\textbf{Stage 3, closed-loop arterial model with a 0D heart model:} To complete the closed-loop cardiovascular
model, systemic and pulmonary venous pathways were incorporated using 3-element Windkessel models,
linking arterial outlets to the atria of the lumped-parameter heart model. Parameters including Windkessel
resistances, compliances, and initial nodal pressures were iteratively adjusted to match clinical data. For each
patient, between 30 and 60 parameters were calibrated until the relative error between simulated and measured
hemodynamic metrics—such as mean, systolic, diastolic pressures, and mean flow—fell below 10\%. Percent error
were calculated using the formula $(H_{i}^{data} - H_{i}^{model})/H_{i}^{data} * 100$, where
$H_{i}$ $\{P_{\textit{mean}}, P_{\textit{systolic}}, P_{\textit{diastolic}}, Q_{\textit{mean}}\}$ after achieving cycle-to-cycle
convergence in the simulations.

\textbf{Model calibration:} Simulated mean flow rates, mean pressures, systolic, diastolic, and pulse pressures were all matched within
10\% of clinical data in the MPA and ascending aorta through an iterative process, as shown in Figure 5.

\begin{figure}[h]
    \centering
    \includegraphics[width=0.4\textwidth]{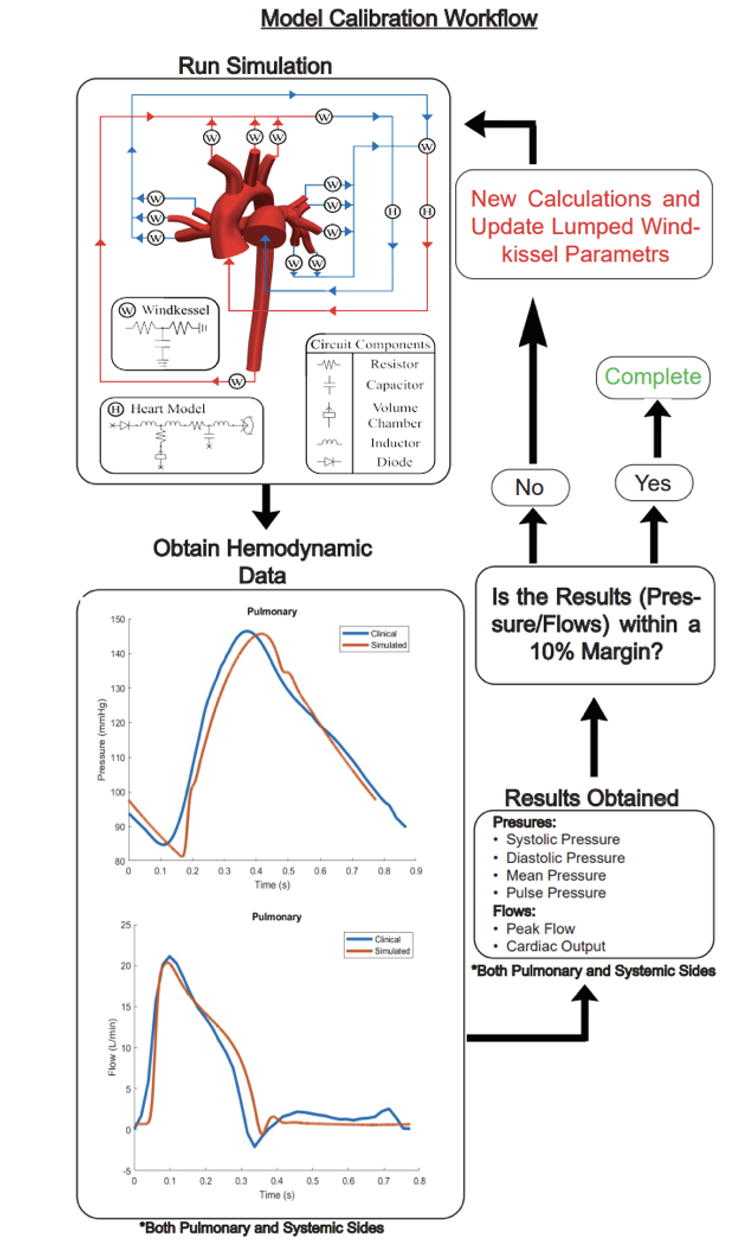}
    \caption{Model calibration workflow for tuning lumped-parameter Windkessel components. Simulations are iteratively updated until simulated pressure and flow waveforms match clinical data within a 10\% margin for both pulmonary and systemic circulations.}
    \label{fig:model_calibration_workflow}
\end{figure}

\textbf{Optimizer:} To streamline the calibration of patient-specific cardiovascular models, we developed a fully automated Python-
based optimizer (Figure 6). This tool interfaces with CRIMSON’s flow solver using subprocess, and processes simulation outputs
with NumPy, pandas, and other standard libraries. It replaces the previously manual, time-consuming process of tuning lumped
Windkessel and heart model parameters. The optimizer iteratively adjusts resistances, compliances, and source pressures across all
Windkessel outlets and heart model inlets. At each iteration, it
compares simulation results to clinical targets—including mean, systolic, diastolic, and pulse pressures,
as well as mean and peak flow rates in the MPA and ascending aorta. The script
evaluates error convergence against a 10\% threshold for all key metrics, recalibrating parameters as
needed until physiological agreement is reached. The calibration cycle uses a feedback-driven approach,
modifying over 40 parameters in the netlist (e.g., Rp, Rd, C values) and automatically adjusting pressure
boundary conditions and source resistances when applicable. This automation reduced the overall
calibration time per patient from 1–2 weeks to just 2–4 days and ensured reproducibility, accuracy, and
scalability of the simulation pipeline. By fully integrating simulation control, data extraction, and
optimization logic, the script lays the groundwork for high-throughput modeling of large clinical cohorts.

\begin{figure}[h]
    \includegraphics[width=0.53\textwidth]{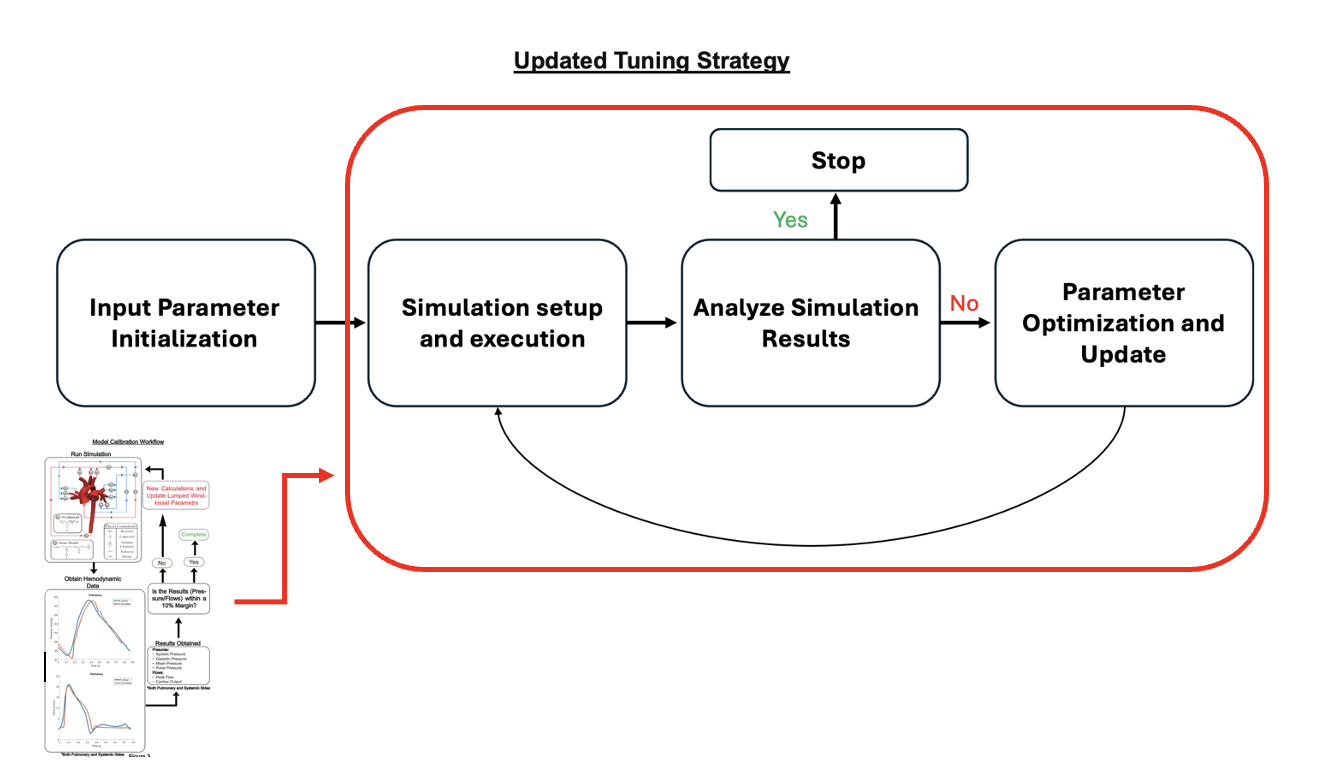}
    \caption{Updated tuning strategy for model calibration. This automated workflow streamlines the iterative process of parameter adjustment by executing input initialization, simulation, analysis, and optimization cycles without manual intervention, continuing until hemodynamic targets are met.}
    \label{fig:updated_tuning_strategy}
\end{figure}

\textbf{Model-derived metrics:} A total of 7 model-derived metrics and parameters were obtained from the high-
resolution arterial models and included in the disease prognosis study. Stiffness derived at five anatomical
locations (AAo, DTA, MPA, LPA, and RPA), MPA-LPA, MPA-RPA, and AAo-DTA pulse wave velocities, and
distribution of central and peripheral pulmonary vasculature resistance and compliance$^{25}$ were evaluated.

\section*{Results}

After simulation results successfully matched patient-specific clinical hemodynamic data within 10\%.
The hemodynamic analyses of four patients at baseline and follow-up demonstrated considerable changes
in MPA stiffness, flow, and pressure dynamics, as derived from both clinical and
simulation data. Table 2 (in the Appendix) shows the model-derived metrics and the clinical derived metrics for each
patient.

In Patient \#1, the MPA stiffness significantly increased from baseline (134,683.6 MPa) to follow-up
(430,298.3 MPa). Correspondingly, to a rise in pulmonary arterial mean pressure (59.4 mmHg to 113.3
mmHg) and pulmonary arterial pulse pressure (30.6 mmHg to 61.7 mmHg) was observed. Pulmonary
vascular resistance index showed a mild increase (16.2 to 18.2 WU$\cdot$m$^2$), and pulmonary arterial
compliance decreased from 8.45 mm$^4\cdot$s$^2$/g to 6.54 mm$^4\cdot$s$^2$/g.

Patient \#2 presented relatively smaller changes in arterial stiffness from baseline (324,985.9 MPa) to
follow-up (350,659.5 MPa), and a reduction in pulmonary arterial mean pressure (82.9 mmHg to 55.0
mmHg) and pulse pressure (66.2 mmHg to 61.8 mmHg). The pulmonary arterial compliance increased
from 3.45 mm$^4\cdot$s$^2$/g to 4.33 mm$^4\cdot$s$^2$/g, indicating improved vascular elasticity.

For Patient \#3, arterial stiffness exhibited a decrease from baseline (100,562.5 MPa) to follow-up
(71,609.6 MPa). Despite the reduction in stiffness, pulmonary vascular resistance significantly increased
(7.3 to 9.1 WU$\cdot$m$^2$). Compliance similarly reduced from 20.49 mm$^4\cdot$s$^2$/g to
16.45 mm$^4\cdot$s$^2$/g, with minimal changes in pulmonary arterial mean and pulse pressures.

Lastly, Patient \#4 demonstrated an decrease in stiffness (190,034.9 MPa to 98,286.2 MPa), along with a
reduction in pulmonary arterial mean pressure (50.1 mmHg to 27.3 mmHg) and pulse pressure (33.3
mmHg to 27.7 mmHg). Pulmonary arterial resistance and compliance showed substantial changes, with
compliance increasing from 7.03 mm$^4\cdot$s$^2$/g to 10.60 mm$^4\cdot$s$^2$/g, reflecting improved
arterial elasticity.

Across all patients, simulated flow and pressure waveforms closely matched the clinical data, validating the accuracy of the modeling approach as seen in Figure 7 in the Appendix. The observed stiffness distributions revealed spatial heterogeneity and patient-specific pulmonary artery remodeling under varying hemodynamic conditions. Importantly, longitudinal changes in model-derived stiffness and compliance consistently tracked directional changes in clinical hemodynamics, even when traditional metrics such as cardiac index remained stable, suggesting that computational metrics may be sensitive to subclinical progression or improvement.

\section*{Discussion}

In Patient \#1, an increase in both MPA systolic and pulse pressures increased RV workload, as demonstrated by the significant increase in RVSWi, which has detrimental effects on long-term RV function. The MPA stiffness, a strong indicator of PAH prognosis, increased by approximately 25\%. The greater the stiffness, the larger the pulse pressures in the pulmonary arteries. The patient experienced a 90.7\% increase in mean pulmonary arterial pressure and a 101.6\% rise in pulse pressure. These changes coincided with a 219.5\% increase in model-derived MPA stiffness and marked increases in pulse wave velocities (PWVs) across both left and right pulmonary arteries (111.4\% and 158.6\%, respectively). In addition, the RV Ejection Fraction declined from 50.0\% to 44.0\%, and RV Mass Index nearly doubled, indicators of deteriorating RV performance, corresponding with elevated RVSWi (+142.8\%) and vascular resistance. Despite only a slight increase in cardiac index (+2.7\%), the data demonstrate a compensatory mechanism that maintains output at the expense of higher vascular load. The model also captured the drop in pulmonary compliance (-22.6\%), consistent with the reduced MPA relative area change (-38.6\%) observed in MRI.

Despite most hemodynamic metrics improving in Patient \#2, a slight 0.5\% increase in MPA stiffness was observed. Changes in pulmonary arterial PWVs, directly affects the arterial stiffness, were observed in both patients. Cardiac indexes remained constant irrespective of changes in disease state, reflecting the system’s ability to adapt and maintain a homeostatic cardiac output at the expense of increasing pressure and cardiac work. The MRI data revealed an increase in RV ejection fraction (from 49.0\% to 60.0\%) and a reduction in RV end-systolic volume index (-24.1\%), both suggestive of improved ventricular function. Meanwhile, catheterization data showed a 33.6\% drop in mean pulmonary arterial pressure and a 28.0\% decrease in vascular resistance, indicating significant hemodynamic relief. These improvements were mirrored by the model-derived drop in pulmonary arterial resistance (-43.9\%) and increase in compliance (+25.7\%). Interestingly, while RV stroke work index decreased by 43.7\%, suggesting reduced RV workload, MPA stiffness increased slightly, indicating localized stiffening not necessarily captured by bulk pressure or volume measurements. The model captured the disease prognosis through increasing in stiffness in the MPA and LPA, alongside minor shifts in PWVs.

Patient \#1's presented a more complex disease, including both PAH and RV dysfunction, in addition to anemia, mild obstructive sleep apnea, and thrombocytopenia. The large age gap between patients could explain the differential outcomes in disease progression.

While there was an improvement in Patient \#3's MPA Stiffness and RVSWi, with decrease of 38.2\% and 28.8\% respectively, we observed no significant alteration in Pulmonary Arterial Systolic Pressures. Additionally, it's worth noting an increase in MPA-LPA PWVs, despite the observed decrease in MPA stiffness. MRI-derived data revealed minimal changes in RV function—RV end-diastolic and end-systolic volume indices remained essentially unchanged, and RV ejection fraction only rose by 1.9\%. However, there were significant increases in MPA systolic and diastolic area indices, resulting in a near doubling of the relative area change (+98\%), a surprising finding that was not reflected in the pressure data. Catheter-based measurements suggested a rise in vascular resistance (from 7.3 to 9.1 WU·m\textsuperscript{2}), which was not fully captured by the model-derived resistance values that remained stable at 0.05 g/(mm\textsuperscript{4}·s). This discrepancy could reflect the model limitations in small vessel remodeling. Additionally, pulmonary arterial compliance decreased (-19.7\%), in alignment with the overall stiffening seen in the LPA and RPA (+59.9\% and +23.8\%, respectively).

Patient \#4 was diagnosed with patent ductus arteriosus (PDA), a congenital cardiac defect characterized by a persistent opening between the aorta and the pulmonary artery, potentially leading to abnormal hemodynamic conditions. The reductions in pulmonary arterial pressures and vascular resistance were mirrored by decreases in MPA stiffness, reinforcing the physiological relevance of the model. MRI-derived data showed a decrease in MPA relative area change (-24.3\%), stroke volume index (-23.1\%), and RV mass index (-25.5\%), suggesting reduced cardiac workload, despite a modest increase in cardiac index (3.7\%) and heart rate (+16\%). These patterns align with improved hemodynamic conditions following PDA closure. Catheter-derived metrics confirmed this trend, showing a 45.5\% reduction in pulmonary arterial mean pressure and a 30.4\% drop in pulmonary vascular resistance index. These changes coincided with reductions in RV stroke work index (-23.1\%) and RV pressure ratios, indicating improved ventricular unloading. Model-derived metrics closely matched these clinical trends. MPA stiffness dropped by 48.3\%, accompanied by decreases in LPA and RPA stiffness (25.0\% and 33.4\%, respectively). Pulmonary arterial resistance also declined by 38.8\%, while compliance increased by 50.9\%, further validating the model's ability to capture pressure–volume relationships. Interestingly, while MPA-RPA PWV decreased substantially (-66.7\%), MPA-LPA PWV remained nearly constant (-4.5\%), indicating asymmetric vascular responses across the pulmonary circuit.

More broadly, across all four patients, strong relationships were noted between clinical (MRI and Cath) and model-derived metrics. For example, increases in pulmonary arterial pressures or vascular resistance were generally associated with corresponding elevations in model-derived stiffness and pulse wave velocities. Similarly, reductions in compliance captured by the model aligned with clinical signs of vascular stiffening. These relationships support the validity of the computational modeling framework in capturing patient-specific hemodynamic trends and underscores its potential to serve as a non-invasive alternative for invasive measurements.

\section*{Limitations}

This study has several limitations. First, we present results from only four patients, all of whom were female. While these cases demonstrate the validity of the multi-scale modeling approach, the limited sample size restricts the generalizability of our findings. Two additional patients were excluded from this analysis because their simulation results had not yet been fully processed at the time of manuscript preparation. Including these cases in future work may provide broader insight into sex-specific remodeling patterns or inter-patient variability.

Second, one of the four patients included in this study had a congenital defect PDA, as mentioned before, introducing additional variability within an already small patient cohort. This heterogeneity limits the ability to generalize the results and weakens the strength of observed correlations. To validate and extend these findings, future studies should include larger and more diverse patient populations.

Third, the current model does not include a detailed representation of ventricular mechanics. The exclusion of a full ventricular model limits our ability to capture the complete hemodynamic state of the heart, including dynamic ventricular–arterial coupling and pressure–volume relationships. This simplification may overlook key interactions between ventricular function and pulmonary hemodynamics, particularly in conditions involving RV remodeling or failure.

\section*{Conclusion}

The findings from this study provide valuable insights into the hemodynamic changes associated with the progression of PAH. By aligning model-derived metrics with clinical hemodynamic data, we demonstrated the ability of patient-specific computational modeling to assess disease severity and progression in pediatric PAH cases. Notably, the model captured key trends in pulmonary arterial stiffness, resistance, compliance, and RV workload that closely mirrored changes observed in MRI- and catheter-based measurements.

Through detailed case studies, we showed how the modeling framework can differentiate between patients with worsening PAH, those exhibiting signs of hemodynamic improvement, and those with congenital anomalies. These individualized assessments underscore the model’s potential to enhance clinical understanding of complex cardiopulmonary conditions and to support longitudinal, non-invasive monitoring of disease progression.

Future work will focus on integrating high-fidelity finite element models of the right ventricle to study RV remodeling in greater detail, building upon the methodology established in our foundational work. Additionally, expanding the patient cohort and incorporating male patients will be essential for validating model generalizability. Further refinements may include incorporating a full ventricular model to improve the physiological fidelity of ventricular–arterial interactions. We also plan to extract additional model-derived metrics, such as central pulmonary arterial resistance and compliance, which may offer deeper insights into disease progression and underlying pathophysiology.
    
\section*{Acknowledgements}

This work was supported by the National Institutes of Health (NIH) under grant U01 HL135842.

\nocite{*}
\bibliographystyle{unsrt}
\bibliography{references}

@article{Simonneau2019,
  author  = {Simonneau, G. and Montani, D. and Celermajer, D. S. and others},
  title   = {Haemodynamic definitions and updated clinical classification of pulmonary hypertension},
  journal = {European Respiratory Journal},
  year    = {2019},
  volume  = {53},
  number  = {1},
  pages   = {1801913},
  doi     = {10.1183/13993003.01913-2018}
}

@article{Humphrey2016,
  author  = {Humphrey, J. D. and Harrison, D. G. and Figueroa, C. A. and Lacolley, P. and Laurent, S.},
  title   = {Central artery stiffness in hypertension and aging},
  journal = {Circulation Research},
  year    = {2016},
  volume  = {118},
  number  = {3},
  pages   = {379--381},
  doi     = {10.1161/CIRCRESAHA.115.307722}
}

@article{VonkNoordegraaf2019,
  author  = {Vonk Noordegraaf, A. and Chin, K. M. and Haddad, F. and others},
  title   = {Pathophysiology of the right ventricle and of the pulmonary circulation in pulmonary hypertension: An update},
  journal = {European Respiratory Journal},
  year    = {2019},
  volume  = {53},
  number  = {1},
  pages   = {1801900},
  doi     = {10.1183/13993003.01900-2018}
}

@article{VonkNoordegraaf2017,
  author  = {Vonk Noordegraaf, A. and Westerhof, B. E. and Westerhof, N.},
  title   = {The relationship between the right ventricle and its load in pulmonary hypertension},
  journal = {Journal of the American College of Cardiology},
  year    = {2017},
  volume  = {69},
  number  = {19},
  pages   = {236--243},
  doi     = {10.1016/j.jacc.2016.10.047}
}

@article{Beghetti2014,
  author  = {Beghetti, M. and Berger, R. M. F.},
  title   = {The challenges in pediatric pulmonary arterial hypertension},
  journal = {European Respiratory Review},
  year    = {2014},
  volume  = {23},
  number  = {134},
  pages   = {498--504},
  doi     = {10.1183/09059180.00007714}
}

@article{TossasBetancourt2022,
  author  = {Tossas-Betancourt, C. and Li, N. Y. and Shavik, S. M. and others},
  title   = {Data-driven computational models of ventricular--arterial hemodynamics in pediatric pulmonary arterial hypertension},
  journal = {Frontiers in Physiology},
  year    = {2022},
  volume  = {13},
  pages   = {958734},
  doi     = {10.3389/fphys.2022.958734}
}

@article{Arthurs2021CRIMSON,
  title   = {CRIMSON: An open-source software framework for cardiovascular integrated modelling and simulation},
  author  = {Arthurs, Christopher J. and et al.},
  journal = {PLOS Computational Biology},
  year    = {2021},
  volume  = {17},
  number  = {5},
  pages   = {e1008881},
  doi     = {10.1371/journal.pcbi.1008881}
}

@article{VignonClementel2010,
  author  = {Vignon-Clementel, I. E. and Figueroa, C. A. and Jansen, K. E. and Taylor, C. A.},
  title   = {Outflow boundary conditions for three-dimensional simulations of non-periodic blood flow and pressure fields in deformable arteries},
  journal = {Computer Methods in Biomechanics and Biomedical Engineering},
  year    = {2010},
  volume  = {13},
  number  = {5},
  pages   = {625--640},
  doi     = {10.1080/10255840903413565}
}

@article{Figueroa2006,
  author  = {Figueroa, C. A. and Vignon-Clementel, I. E. and Jansen, K. E. and Hughes, T. J. R. and Taylor, C. A.},
  title   = {A coupled momentum method for modeling blood flow in three-dimensional deformable arteries},
  journal = {Computer Methods in Applied Mechanics and Engineering},
  year    = {2006},
  volume  = {195},
  number  = {41--43},
  pages   = {5685--5706},
  doi     = {10.1016/j.cma.2005.11.011}
}

@article{Hirai1989,
  author  = {Hirai, T. and Sasayama, S. and Kawasaki, T. and Yagi, S.},
  title   = {Stiffness of systemic arteries in patients with myocardial infarction},
  journal = {Circulation},
  year    = {1989},
  volume  = {80},
  number  = {1},
  pages   = {78--86},
  doi     = {10.1161/01.CIR.80.1.78}
}

@article{SilvaVieira2018,
  author  = {Silva Vieira, M. and Arthurs, C. J. and Hussain, T. and Razavi, R. and Figueroa, C. A.},
  title   = {Patient-specific modeling of right coronary circulation vulnerability post-liver transplant in Alagille’s syndrome},
  journal = {PLOS ONE},
  year    = {2018},
  volume  = {13},
  number  = {11},
  pages   = {e0205829},
  doi     = {10.1371/journal.pone.0205829}
}

@article{Roccabianca2014,
  author  = {Roccabianca, S. and Figueroa, C. A. and Tellides, G. and Humphrey, J. D.},
  title   = {Quantification of regional differences in aortic stiffness in the aging human},
  journal = {Journal of the Mechanical Behavior of Biomedical Materials},
  year    = {2014},
  volume  = {29},
  pages   = {618--634},
  doi     = {10.1016/j.jmbbm.2013.01.026}
}

@article{Li2012,
  author  = {Li, N. and Zhang, S. and Hou, J. and Jiang, K.-K. and Yu, B.},
  title   = {Assessment of pulmonary artery morphology by optical coherence tomography},
  journal = {Heart, Lung and Circulation},
  year    = {2012},
  volume  = {21},
  number  = {12},
  pages   = {778--781},
  doi     = {10.1016/j.hlc.2012.07.014}
}

@article{Xiao2013,
  author  = {Xiao, N. and Humphrey, J. D. and Figueroa, C. A.},
  title   = {Multi-scale computational model of three-dimensional hemodynamics within a deformable full-body arterial network},
  journal = {Journal of Computational Physics},
  year    = {2013},
  volume  = {244},
  pages   = {22--40},
  doi     = {10.1016/j.jcp.2012.09.016}
}

@article{Lau2015,
  author  = {Lau, K. D. and Figueroa, C. A.},
  title   = {Simulation of short-term pressure regulation during the tilt test in a coupled 3D--0D closed-loop model of the circulation},
  journal = {Biomechanics and Modeling in Mechanobiology},
  year    = {2015},
  volume  = {14},
  number  = {4},
  pages   = {915--929},
  doi     = {10.1007/s10237-014-0645-x}
}

@article{Arthurs2017,
  author  = {Arthurs, C. J. and Lau, K. D. and Girka, R. G. and Figueroa, C. A.},
  title   = {Reproducing patient-specific hemodynamics in the Blalock--Taussig circulation using a flexible multi-domain simulation framework},
  journal = {Frontiers in Pediatrics},
  year    = {2017},
  volume  = {5},
  pages   = {5},
  doi     = {10.3389/fped.2017.00078}
}

@article{Kim2009,
  author  = {Kim, H. J. and Vignon-Clementel, I. E. and Figueroa, C. A. and others},
  title   = {On coupling a lumped parameter heart model and a three-dimensional finite element aorta model},
  journal = {Annals of Biomedical Engineering},
  year    = {2009},
  volume  = {37},
  number  = {11},
  pages   = {2153--2169},
  doi     = {10.1007/s10439-009-9760-8}
}

@article{Sankaran2012,
  author  = {Sankaran, S. and Esmaily Moghadam, M. and Kahn, A. M. and Tseng, E. E. and Guccione, J. M. and Marsden, A. L.},
  title   = {Patient-specific multiscale modeling of blood flow for coronary artery bypass graft surgery},
  journal = {Annals of Biomedical Engineering},
  year    = {2012},
  volume  = {40},
  number  = {10},
  pages   = {2228--2242},
  doi     = {10.1007/s10439-012-0579-3}
}

@article{Marsden2013,
  author  = {Marsden, A. L.},
  title   = {Simulation based planning of surgical interventions in pediatric cardiology},
  journal = {Physics of Fluids},
  year    = {2013},
  volume  = {25},
  number  = {10},
  pages   = {101303},
  doi     = {10.1063/1.4825031}
}

@article{Arthurs2016,
  author  = {Arthurs, C. J. and Lau, K. D. and Asrress, K. N. and Redwood, S. R. and Figueroa, C. A.},
  title   = {A mathematical model of coronary blood flow control: Simulation of patient-specific three-dimensional hemodynamics during exercise},
  journal = {American Journal of Physiology - Heart and Circulatory Physiology},
  year    = {2016},
  volume  = {310},
  number  = {9},
  pages   = {H1152--H1162},
  doi     = {10.1152/ajpheart.00517.2015}
}

@article{Mynard2011,
  author  = {Mynard, J. P. and Davidson, M. R. and Penny, D. J. and Smolich, J. J.},
  title   = {A simple, versatile valve model for use in lumped parameter and one-dimensional cardiovascular models},
  journal = {International Journal for Numerical Methods in Biomedical Engineering},
  year    = {2011},
  volume  = {28},
  number  = {6--7},
  pages   = {626--641},
  doi     = {10.1002/cnm.1466}
}

@article{Ahmed2021,
  author  = {Ahmed, Y. and Tossas-Betancourt, C. and van Bakel, P. A. and others},
  title   = {Interventional planning for endovascular revision of a lateral tunnel Fontan: A patient-specific computational analysis},
  journal = {Frontiers in Physiology},
  year    = {2021},
  volume  = {12},
  pages   = {718254},
  doi     = {10.3389/fphys.2021.718254}
}

@article{Xiao2013Comparison,
  author  = {Xiao, N. and Alastruey, J. and Figueroa, C. A.},
  title   = {A systematic comparison between 1-D and 3-D hemodynamics in compliant arterial models},
  journal = {International Journal for Numerical Methods in Biomedical Engineering},
  year    = {2013},
  volume  = {30},
  number  = {2},
  pages   = {204--231},
  doi     = {10.1002/cnm.2598}
}

@article{Alastruey2016,
  author  = {Alastruey, J. and Xiao, N. and Fok, H. and Schaeffer, T. and Figueroa, C. A.},
  title   = {On the impact of modelling assumptions in multi-scale, subject-specific models of aortic haemodynamics},
  journal = {Journal of the Royal Society Interface},
  year    = {2016},
  volume  = {13},
  number  = {119},
  pages   = {20160073},
  doi     = {10.1098/rsif.2016.0073}
}

@article{Cuomo2019,
  author  = {Cuomo, F. and Ferruzzi, J. and Agarwal, P. and others},
  title   = {Sex-dependent differences in central artery haemodynamics in normal and fibulin-5 deficient mice},
  journal = {Proceedings of the Royal Society A},
  year    = {2019},
  volume  = {475},
  number  = {2221},
  pages   = {20180076},
  doi     = {10.1098/rspa.2018.0076}
}

\clearpage
\appendix
\onecolumn
\section*{Appendix}

\begin{table}[H]
\centering
\includegraphics[width=0.98\textwidth]{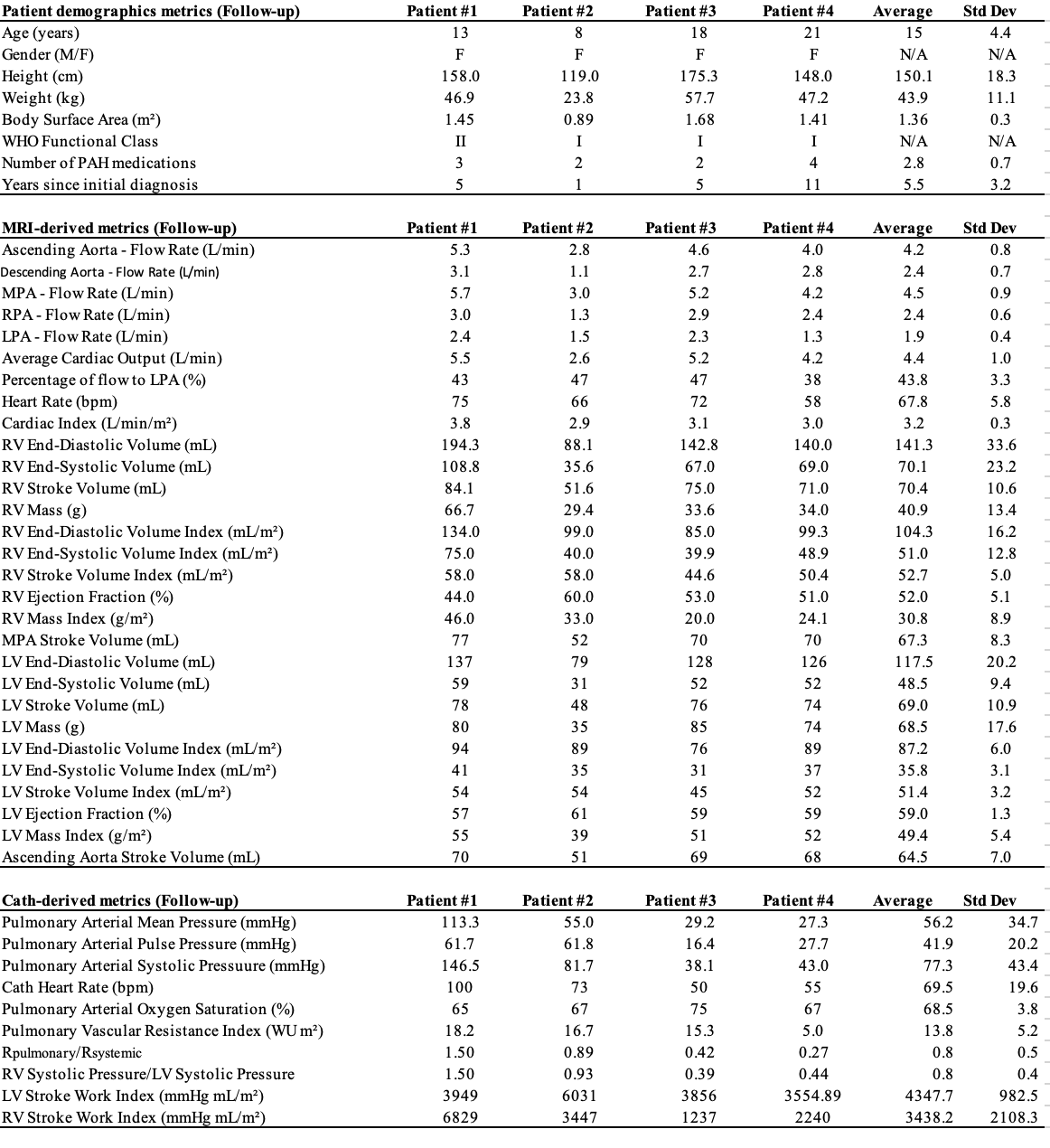}
\caption{Summary of patient demographics and follow-up hemodynamic metrics. Includes MRI-derived and catheter-derived metrics across four female PAH patients, highlighting inter-patient variability and averaged values with standard deviations.}
\label{tab:clinical_metrics}
\end{table}

\clearpage
\onecolumn

\captionsetup{type=table}

\begin{subtable}{0.95\textwidth}
    \centering
    \includegraphics[width=\linewidth]{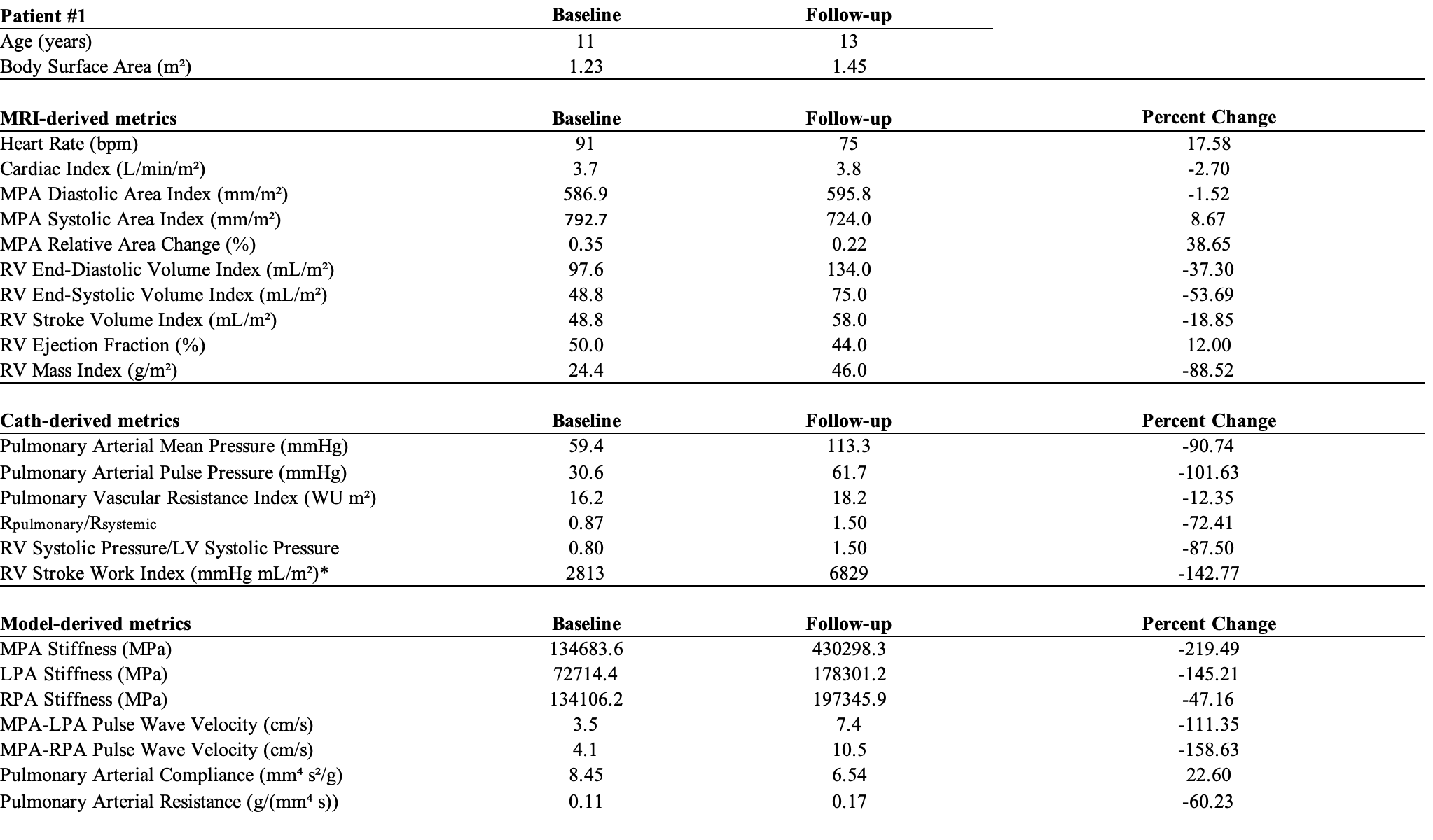}
    \caption*{Patient \#1}
\end{subtable}

\vspace{1em}

\begin{subtable}{0.95\textwidth}
    \centering
    \includegraphics[width=\linewidth]{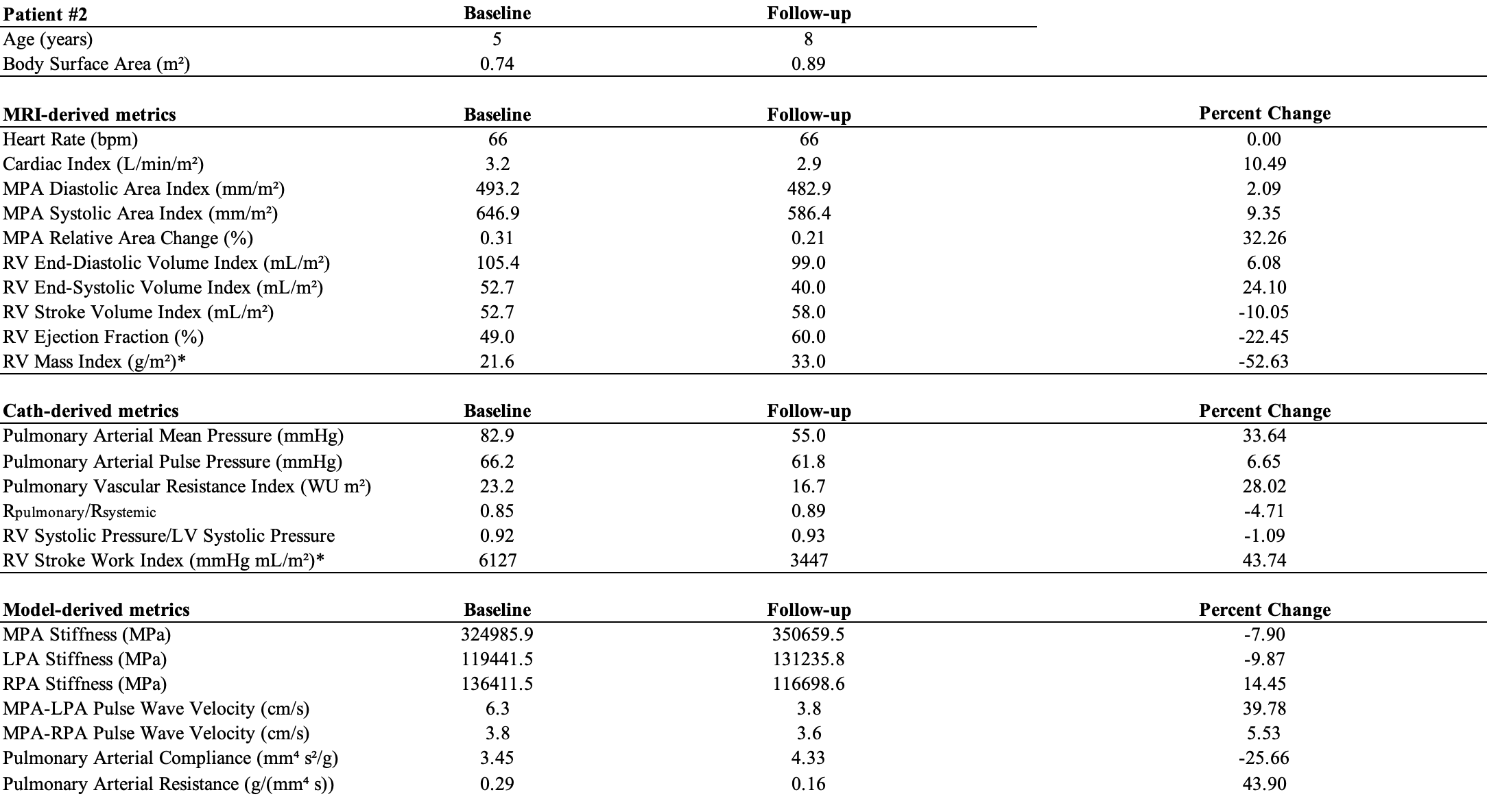}
    \caption*{Patient \#2}
\end{subtable}

\clearpage  

\begin{subtable}{0.95\textwidth}
    \centering
    \includegraphics[width=\linewidth]{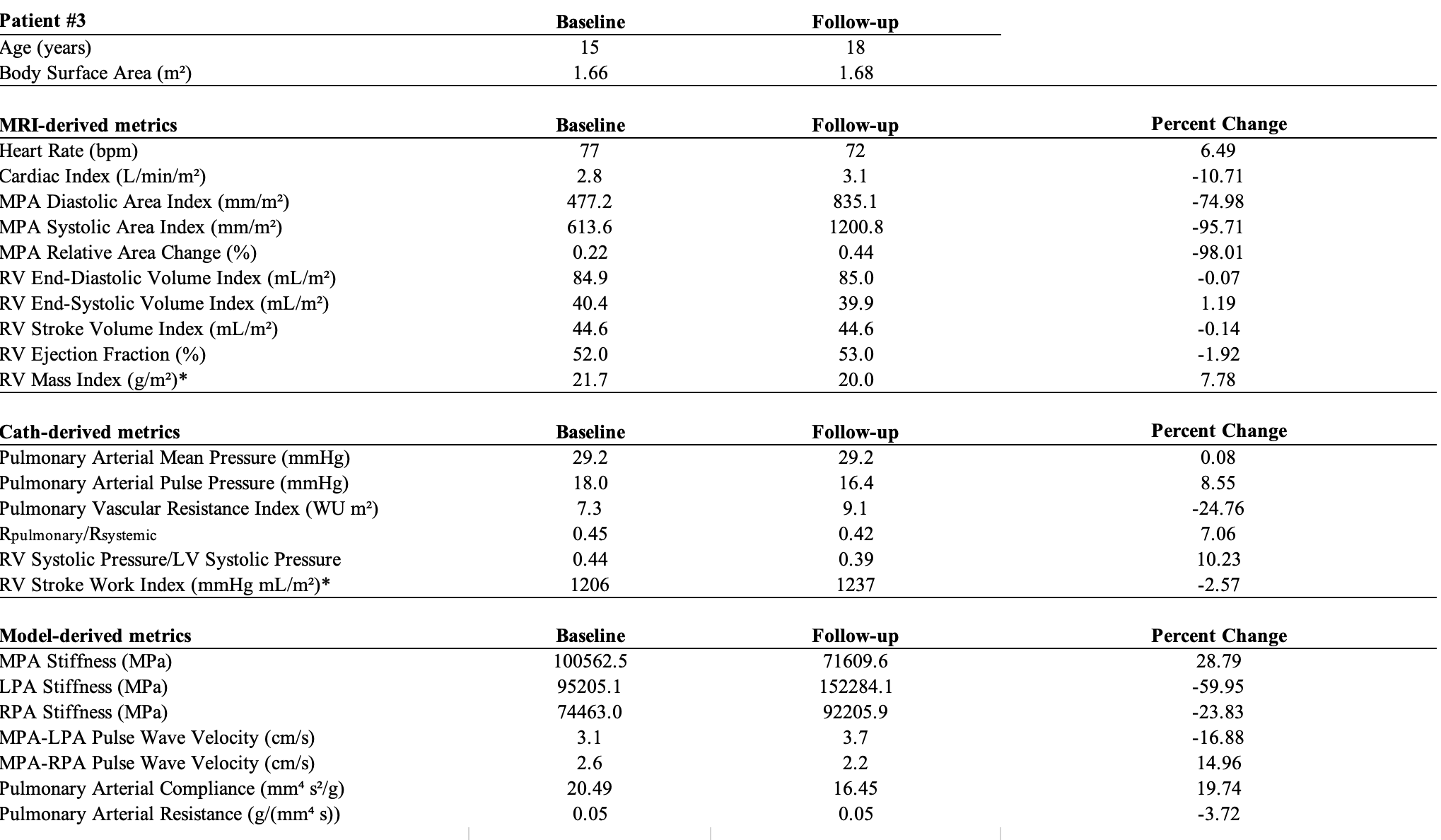}
    \caption*{Patient \#3}
\end{subtable}

\vspace{1em}

\begin{subtable}{0.95\textwidth}
    \centering
    \includegraphics[width=\linewidth]{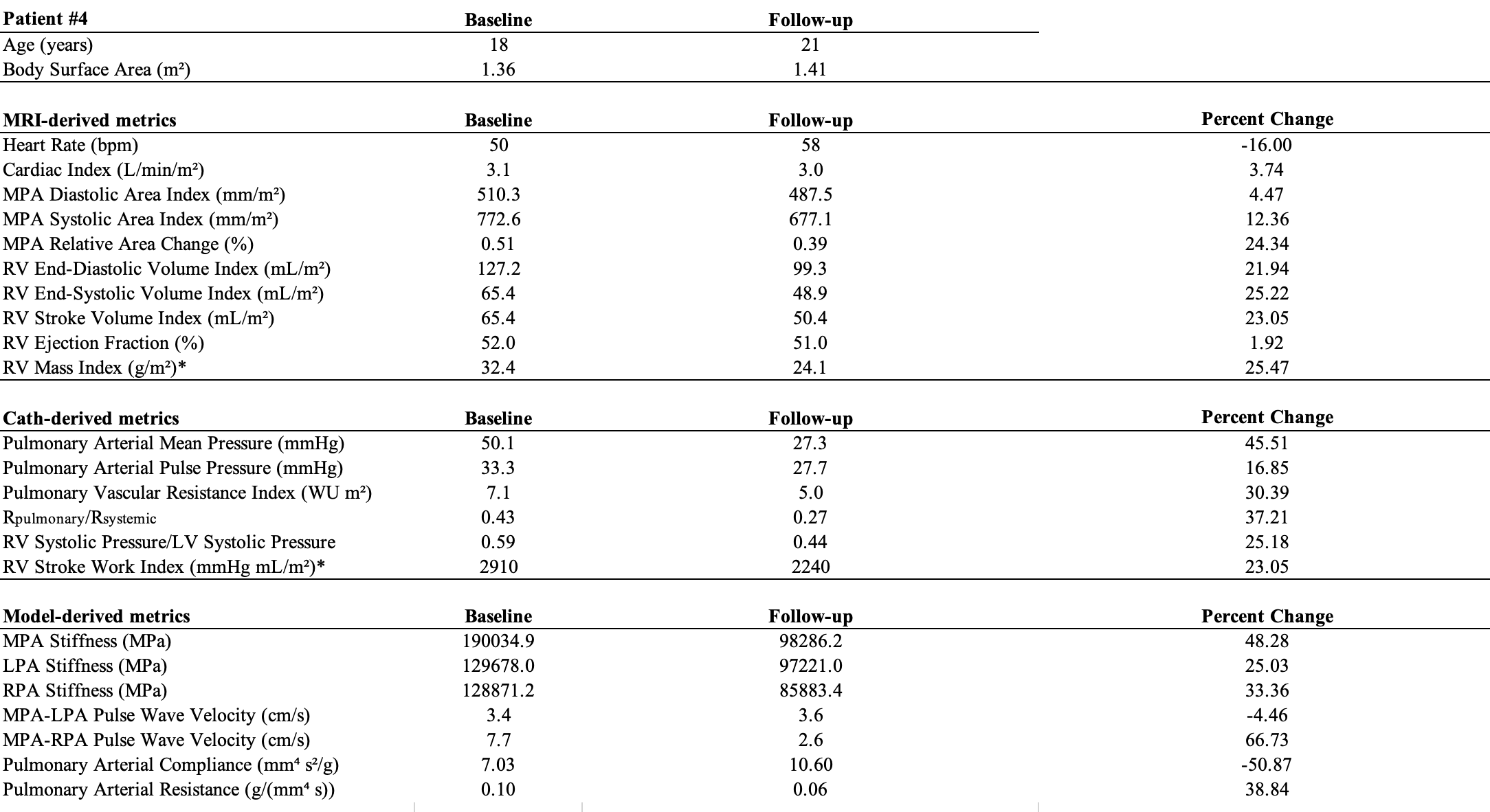}
    \caption*{Patient \#4}
\end{subtable}
\setcounter{table}{1}
\captionof{table}{Baseline and follow-up clinical and model-derived metrics for all four pediatric PAH patients. Metrics include MRI-derived, catheter-derived, and model-derived parameters, along with percent change between baseline and follow-up.}
\label{tab:table2}

\begin{figure*}[t]
    \centering
    \includegraphics[width=\textwidth]{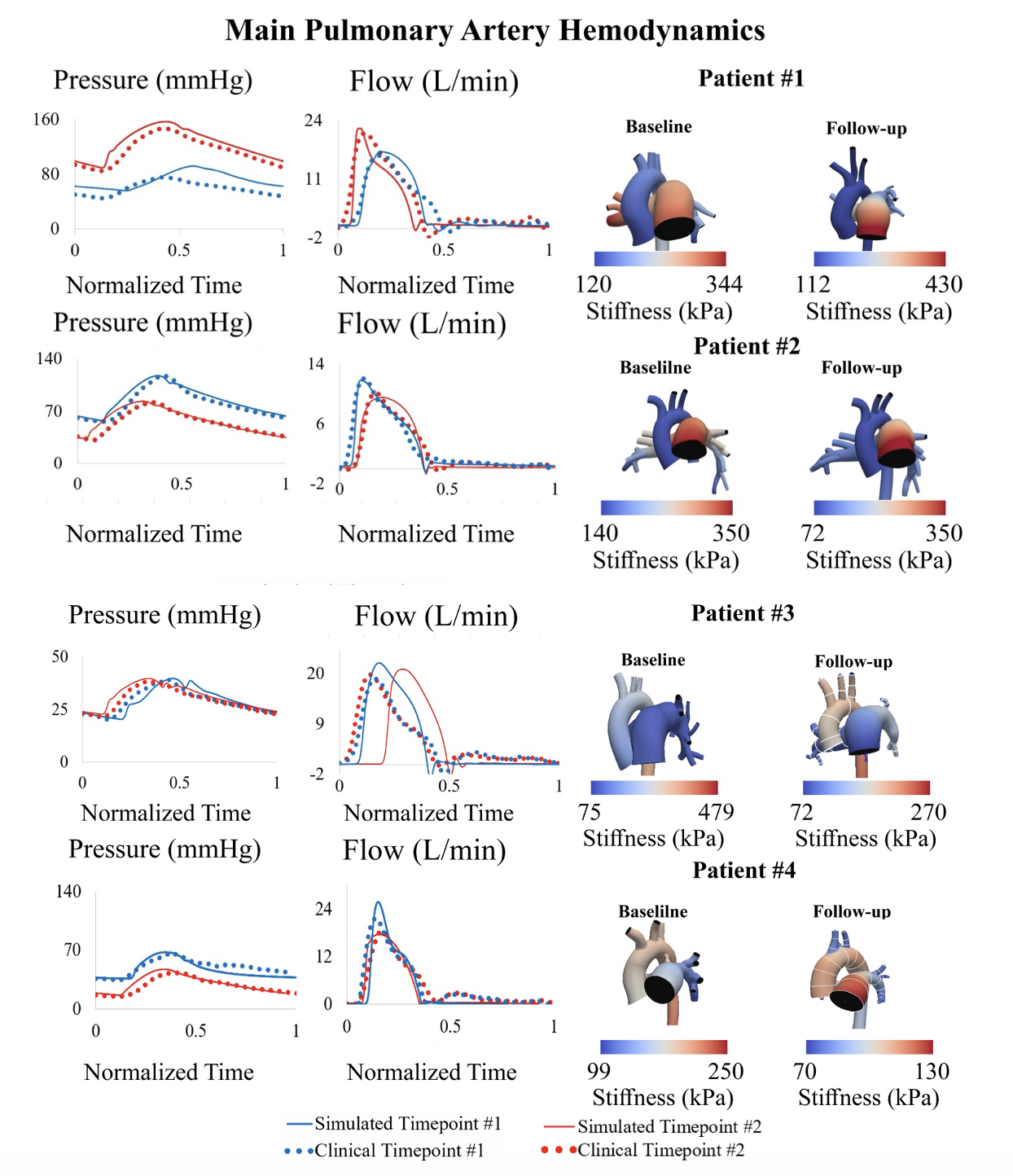}
    \caption{Comparison of simulated and clinical main pulmonary artery (MPA) pressure and flow waveforms at two timepoints (baseline and follow-up) for four PAH patients. Right panels show spatial distributions of Aorta and MPA stiffness derived from imaging and modeling, highlighting individualized progression of vascular stiffness over time.}
    \label{fig:mpa_hemodynamics}
\end{figure*}

\end{document}